\documentclass[aps,prl, twocolumn,a4paper,showpacs,floatfix,superscriptaddress]{revtex4}
\usepackage{amssymb}
\usepackage{amsfonts}
\usepackage{amsmath,bm}
\usepackage{graphics}
\usepackage{color,times}
\usepackage{sidecap}
\usepackage{epsfig}
\usepackage{amssymb,amsfonts,amsmath}

\usepackage{color}
\begin{document}

%%%%%%%%%%%%%%%%%%%%%%%%%%%%%%

%% For titles, only capitalize the first letter
%% \title{Almost sharp fronts for the surface quasi-geostrophic equation}

\title{Heterogeneous resource allocation can change social hierarchy in public goods games.}

%% Enter authors via the \author command.  
%% Use \affil to define affiliations.
%% (Leave no spaces between author name and \affil command)

%% Note that the \thanks{} command has been disabled in favor of
%% a generic, reserved space for PNAS publication footnotes.
\author{Sandro Meloni}
\affiliation{Institute for Biocomputation and Physics of Complex Systems (BIFI), University of Zaragoza, Zaragoza 50009, Spain}
\affiliation{Department of Theoretical Physics, University of Zaragoza, Zaragoza 50009, Spain}

\author{Cheng-Yi Xia}
\affiliation{Tianjin Key Laboratory of Computer Vision and System (Ministry of Education) and Key Laboratory of Intelligence Computing and Novel Software Technology, Tianjin University of Technology, Tianjin 300384,P.R.China}

\author{Yamir Moreno}
\affiliation{Institute for Biocomputation and Physics of Complex Systems (BIFI), University of Zaragoza, Zaragoza 50009, Spain}
\affiliation{Department of Theoretical Physics, University of Zaragoza, Zaragoza 50009, Spain}
\affiliation{Complex Networks and Systems Lagrange Lab, Institute for Scientific Interchange, Turin, Italy}

%%%%%%%%%%%%%%%%%%%%%%%%%%%%%%%%%%%%%%%%%%%%%%%%%%%%%%%%%%%%%%%%
%\begin{article}
\date{\today}

\begin{abstract} 
Public Goods Games represent one of the most useful tools to study group interactions between individuals. However, even if they could provide an explanation for the emergence and stability of cooperation in modern societies, they are not able to reproduce some key features observed in social and economical interactions. The typical shape of wealth distribution -- known as Pareto Law -- and the microscopic organization of wealth production are two of them. Here, we introduce a modification to the classical formulation of Public Goods Games that allows for the emergence of both of these features from first principles. Unlike traditional Public Goods Games on networks, where players contribute equally to all the games in which they participate, we allow individuals to redistribute their contribution according to what they earned in previous rounds. Results from numerical simulations show that not only a Pareto distribution for  the payoffs naturally emerges but also that if players don't invest enough in one round they can act as defectors even if they are formally cooperators. Finally, we also show that the players self-organize in a very productive backbone that  covers almost perfectly the minimum spanning tree of the underlying interaction network. Our results not only give an explanation for the presence of the wealth heterogeneity observed in real data but also points to a conceptual change regarding how cooperation is defined in collective dilemmas. 

\end{abstract}

\maketitle

%% The first letter of the article should be drop cap: \dropcap{}
One of the key elements of human and animal societies is the interaction between groups of individuals to achieve a common goal. The study of cooperation and coordination between individuals has always attracted the attention of scientists from very different fields, ranging from biology\cite{maynard} and sociology\cite{fehr07,gintis03} to economy\cite{camerer,nunn}. On the theoretical side, scientists have tackled this problem using the tools offered by evolutionary game theory\cite{axelrod81,smith82,gintis00,nowak06book}, using among others, Public Goods Games (PGG)\cite{hardin68,sigmund,archetti12,perc13}. Mathematically, PGG are usually represented as the N-person version of the prisoner's dilemma\cite{szabo07,santos05,gardenes07,poncela07} where individuals can decide to contribute (cooperate) or not (defect) to the creation of the public goods. The added value generated by the public goods is modeled by a synergy factor $r$ that multiplies the collected investments resulting in a benefit that is divided equally between all the participants of the group. Then,  to mimic the effect of evolution on the two strategies an evolutionary rule is applied to all players simultaneously\cite{axelrod81,nowak06book,hofbauer88,hofbauer03}.  Despite of its simplicity, this representation shows a very rich behavior and demonstrated itself able to reproduce important traits of real world societies. 

\begin{figure}
\includegraphics[width=\columnwidth]{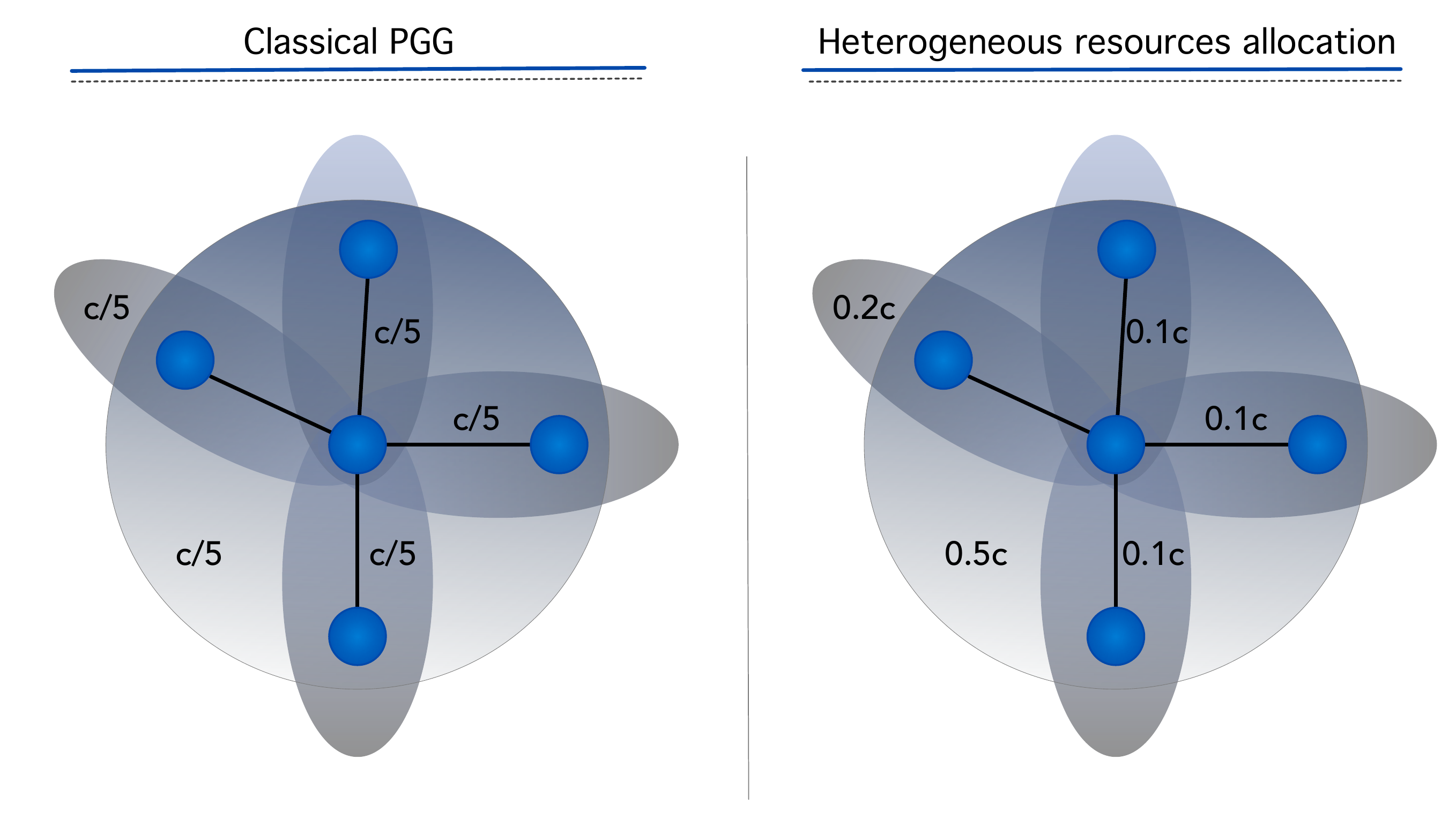}
\caption{Schematic representation of public goods games on networks. In the classical framework (left panel) an individual with four neighbors participates in five different games: the one centered on herself and other four centered on her neighbors and divide her contribution $c$ evenly. In our model (right panel), contributions can be assigned unevenly according to the payoff obtained in the previous round, allocating more resources in more profitable games with a higher probability.}
\label{fig0}
\end{figure}

\section{Results}
\begin{figure}
\includegraphics[width=\columnwidth]{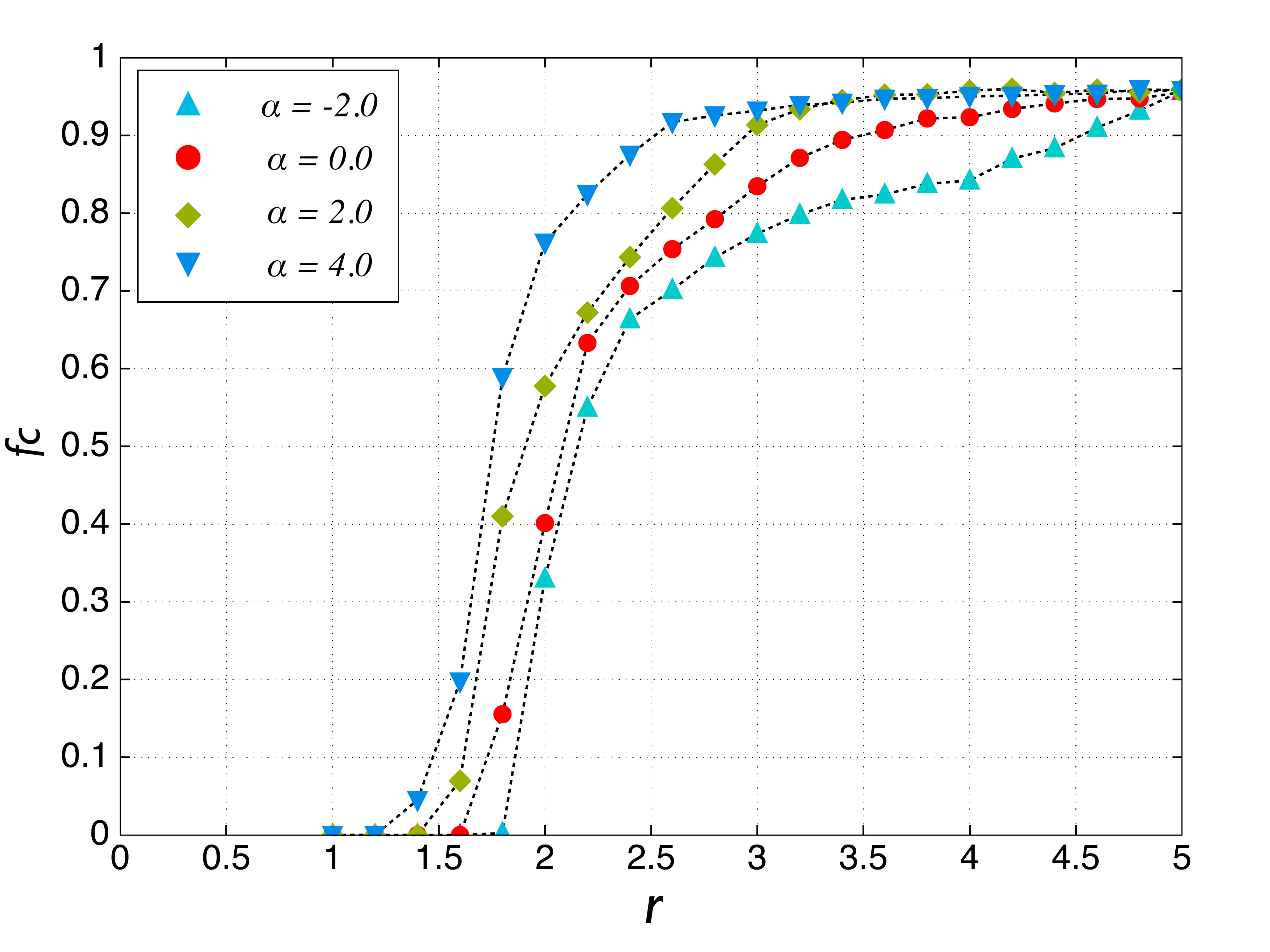}
\caption{Fraction of cooperators $f_c$ as a function of the enhancement factor $r$ for different values of $\alpha$. Results show that there is an increase in the levels of cooperation for highly heterogenous resource allocation (positive values of $\alpha$, see legend). The interaction topology is an uncorrelated scale-free network\cite{Catanzaro2005} with exponent $\gamma = 2.5$ and $N= 10^4$ nodes.  Each point represents an average over at least $500$ runs with randomly chosen initial conditions.}
\label{fig1}
\end{figure}

Recently the search for more realistic models led to the formulation of several modifications of the traditional setup of the PGG. In this direction, one of the first steps has been the introduction of a structure in the population to take into account  the complex interaction patterns present in real societies\cite{boccaletti06,santos08,killingback06}. Simultaneously, other efforts have been put in mimicing realistic traits of our societies like reputation\cite{fehr04}, reward\cite{rand09,szolnoki10}  and punishment mechanisms\cite{fehr00,sigmund07,helbing10a,helbing10b,perc12}, human mobility\cite{cardillo12,helbing09,jiang10,yang10,roca11,cong12,xia12,chen12} and different types of social heterogeneity\cite{peng10,lei10,zhang10,yang12,perc07,shi10,zhang_10,gao10,vukov11,cao_10,zhang_12,kun13}. Heterogeneity seems to play a fundamental role in cooperative behavior -although some results question the role of network heterogeneity \cite{pnas_gracia,sr_gracia}-, and several works have dealt with the effects of allowing an uneven distribution of players' resources\cite{gao10,vukov11,cao_10,zhang_12,kun13}. However, even if the latter studies have helped to understand the emergence of cooperation in large groups, key aspects of the organization of human societies and markets still remain unexplained; a relevant example being the typical wealth distribution observed in economic systems, which cannot be obtained within the formalism of classical PGG.

Here, we address the previous shortcomings and consider a modification of the classical N-person prisoner's dilemma on networks. In our model, players are allowed to distribute their investments unevenly, allocating more resources to profitable games and less in unfavorable ones. In the classical formulation of PGG on networks each neighborhood  is considered as a  group and individuals participate in different groups according to the number of their neighbors. Thus a player with $m$ neighbors will contribute to $m+1$ distinct games (the one centered in her and the $m$ centered on her neighbors) and will divide equally her capital $c$ between all the groups. In our setting, players decide how to distribute their capital according to what they earned in previous rounds (Fig.~\ref{fig0}). We employ a simple distribution function with a parameter $\alpha$ allowing a linear $\alpha=1.0$ or non-linear $\alpha>1.0$  allocation of the resources; $\alpha=0$ recovers the classical formulation with equal investments in all the games (see \textit{Methods} for model's details). 

%% Enter the text of your article beginning here and ending before
%% \begin{acknowledgements}
%% Section head commands for your reference:

Computer simulations of our model for different values of the  allocation parameter $\alpha$ show that when individuals are allowed to distribute their investments unevenly, an increase in the cooperation level is observed with a shift of the critical synergy parameter $r_c$ to lower values (Fig.~\ref{fig1}) with respect to the static allocation scheme $(\alpha = 0)$. The increase in cooperation is more marked for larger values of $\alpha$ while negative values of the parameter --  i.e. invest more in less remunerative games -- lead to a substantial decrease of both the critical synergy factor $r_c$ and the levels of cooperation. Even though these results are consistent with previous studies on similar models\cite{gao10,vukov11,cao_10,zhang_12,kun13} the mechanisms behind this increase and their consequences  on the  organization of the system are still unclear. 

\subsection{Microscopic organization}
\begin{figure*}
\includegraphics[width=0.8\textwidth]{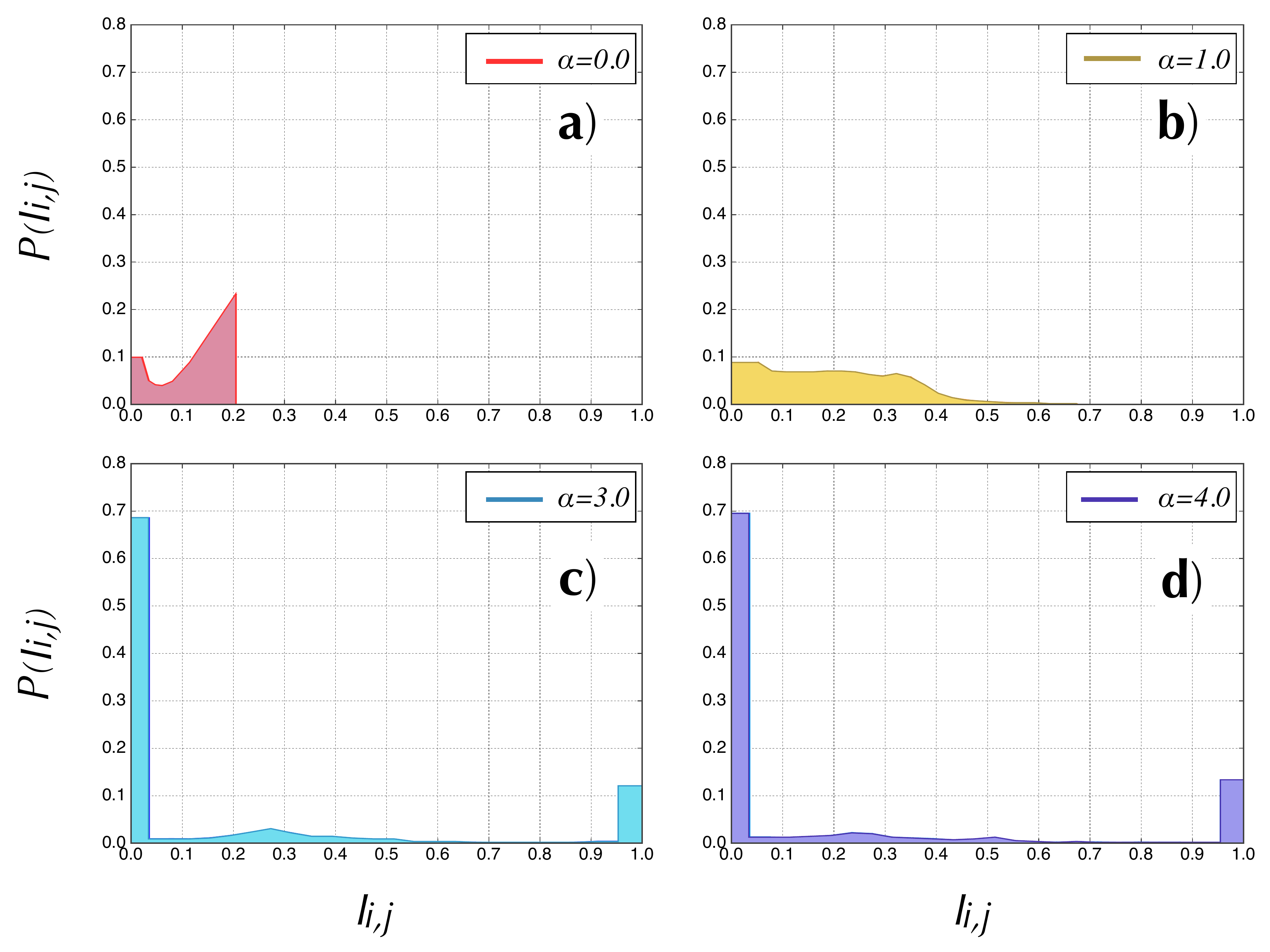}
\caption{Distribution of the investments $I_{x,y}$ over network's links for different values of the resource allocation parameter $\alpha$ at the steady state. A radical change in the distribution is observed from the static case $\alpha = 0$ (panel a), in which the investments follows the degree distribution of the underlying network, to the profit-driven case $\alpha> 0$ (panels b,c and d) in which the invested quantity in a game is related to the previously earned payoff in that game. For non-linear resource allocation $\alpha > 1.0$ (panels c and d) a two-peaked distribution appears where players decide to put almost the totality of their resources in one game and invest a minimal quantity in other games. The substrate topology is an uncorrelated  scale-free network\cite{Catanzaro2005} with exponent $\gamma = 2.5$ and $N= 10^4$ nodes. The synergy factor $r$ is set to $r=4.0$ and $c=1$.}
\label{fig2}
\end{figure*}

To address these questions, we focus on the region where cooperation dominates ($r>3.5$) and look how individuals distribute their investments. Fig.~\ref{fig2} depicts the distribution of the fraction of total investments $I_{i,j}$ for all the players (see \textit{Methods} for details) once the system reached a stationary state. For the static resource allocation  $(\alpha = 0)$ the investment distribution clearly follows the degree distribution of the underlying social graph as players only can distribute their contribution evenly between all the games in which they participate. The picture totally changes even when we consider  a linear allocation of the investments.  As $\alpha$ increases from $0$, the investment distribution rapidly become more heterogenous. Beyond $\alpha=2$, two large peaks centered respectively at very large $( > 0.95)$ and very small (< 0.05) values of $I_{i,j}$ appear.

\begin{figure*}
\includegraphics[width=0.8\textwidth]{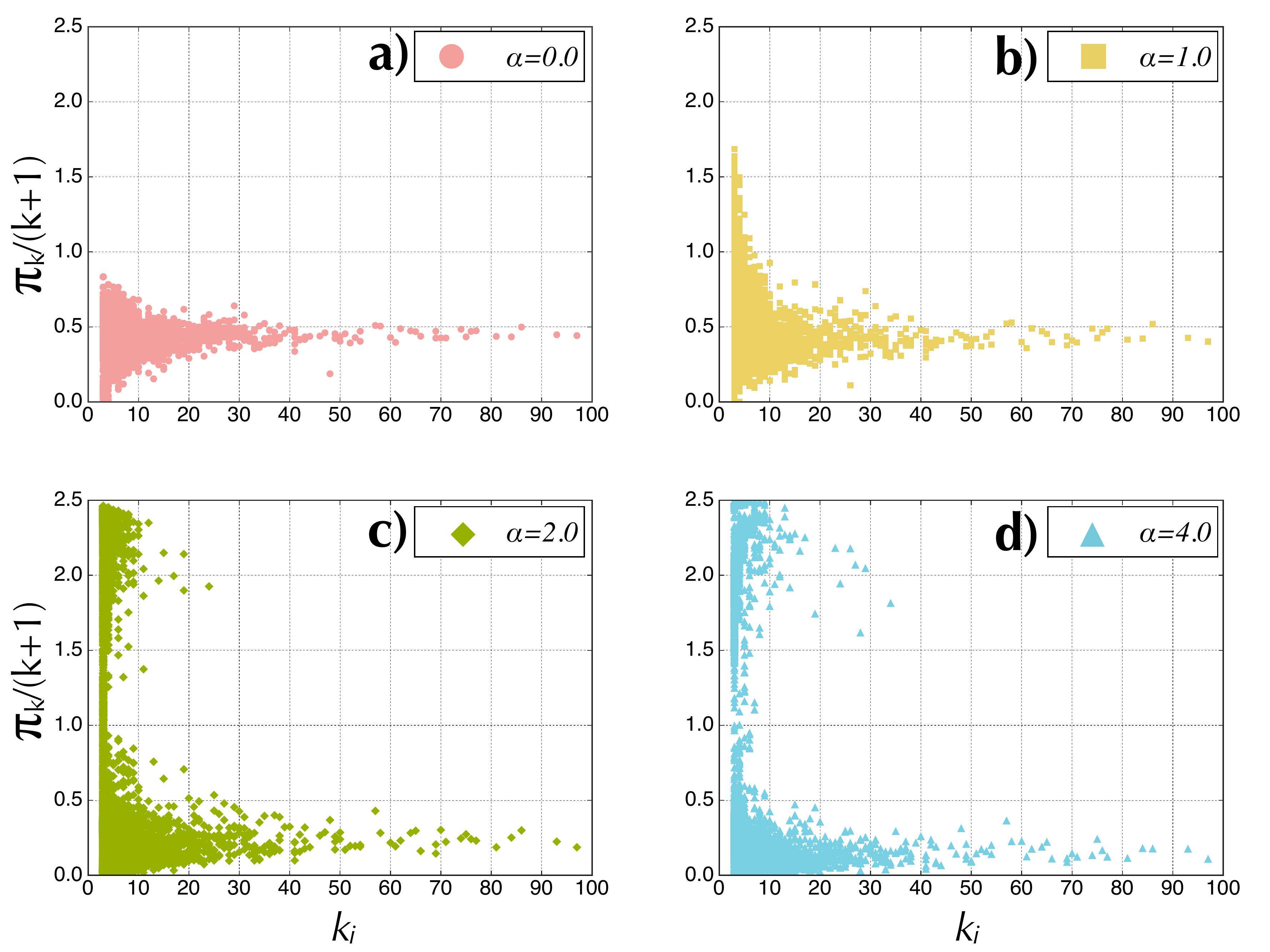}
\caption{Normalized payoff $\Pi_k/(k+1)$ obtained in the game centered at node $i$ as a function of the degree $k$ for different values of the parameter $\alpha$.  For the static case, $\alpha=0$ (panel a), a low dispersion around a mean payoff value is observed for all degrees $k$. Increasing $\alpha$ leads first to an increase in the normalized payoff and its dispersion, most notably for small degrees. A further increase of $\alpha$ produces a second cloud of points localized at the maximal contribution (e.g., c(r-1)) and low degrees (panels b through d). The other parameters are the same as in Fig.~\ref{fig2}. }
\label{fig3}
\end{figure*}

Results shown in Fig.~\ref{fig2} suggest that, once players are free to allocate their resources, a very peculiar organization emerges. The peak for large values of the investments indicates that most of the players (almost the totality for $\alpha \ge 3.0$) allocate the majority of their resources in only one market -- the most profitable one -- while they distribute evenly between all the other games the remaining part of their capital creating the peak for small values of $I_{i,j}$. The previous findings could explain the observed increase in players' cooperation\cite{gao10,vukov11,cao_10,zhang_12,kun13}, but it is not the only consequence of the observed investment distribution. Indeed, an established result in public goods games is that the most connected nodes -- the hubs -- are responsible for the emergence of cooperation and for the production of the majority of the payoff. However, the results in Fig.~\ref{fig3} depict a totally different scenario. If we consider the total normalized payoff produced in games centered on nodes of degree $k$, $\Pi_k/(k+1)$, we find that, in the classical case, it is distributed almost homogeneously among all the degrees, with a mean value around $0.15$  (Fig.~\ref{fig3}a). However, for $\alpha > 0$ strong differences arise. As $\alpha$ increases, the distribution of the payoff for games taking place on low degree nodes starts to become more heterogenous until, finally, for $\alpha = 3.0$ (Fig.~\ref{fig3}d) a large number of games produce very high payoffs (note that the maximum payoff does not depend on $k$, as $\text{max}\{\Pi_k/(k+1)\}=c(r-1)$). This means that all the players invested all their contributions in that game. Moreover, with the increase of $\alpha$ the average payoff produced in the hubs decreases substantially. 

These results, also at the light of the investments' distribution (Fig~\ref{fig2}), indicate a radical change in the social structure of the system. While in classical PGG, hubs represent the driving force and the centers where the majority of the wealth is produced, in our model, games on poorly connected nodes are responsible for the creation of the largest part of the public goods. Specifically, our results demonstrate that individuals self-organize in a large number of small sized clusters formed exclusively by cooperators where all the players invest almost their entire capital. This paradigm shift also has other interesting consequences that will be discussed in the next  sections. 

\subsection{Wealth distribution} One of the criticisms to the classical N-persons' prisoner's dilemma (also on heterogeneous social structures) is that it fails to reproduce the wealth distribution observed in real economic systems -- the so-called Pareto principle\cite{pareto} -- where the $80\%$ of the total wealth is generated by the $20\%$ of the population. This is mainly due to the fact that, even if games centered on hubs provide higher payoffs because many players participate in them, they are only a small fraction of the population (surely much less than the $20\%$) and are not able to significantly change the overall wealth distribution. As in our model cooperators tend to form small but very productive clusters it is interesting to look at the wealth distribution produced for different values of $\alpha$.

\begin{figure}
\includegraphics[width=\columnwidth]{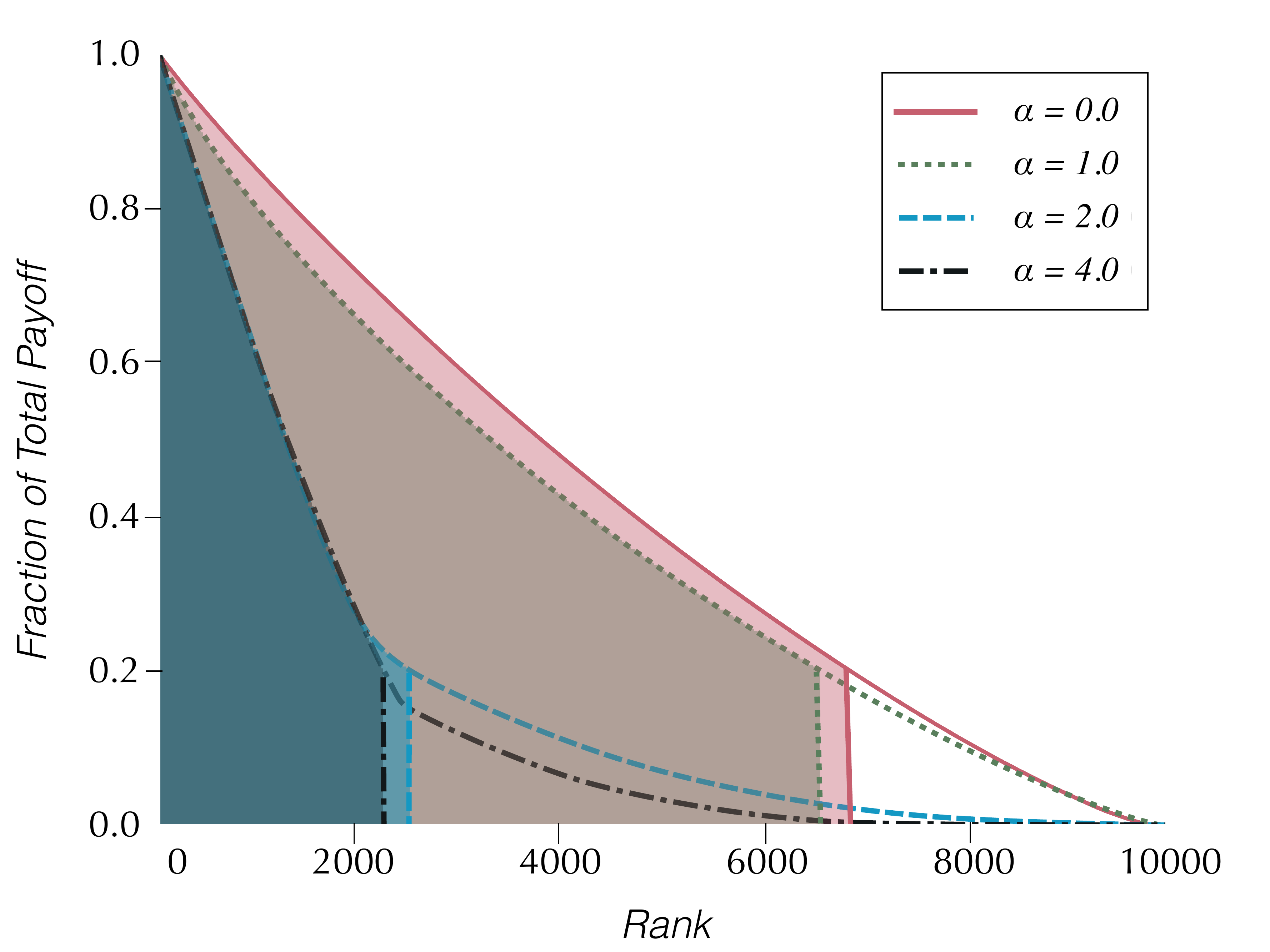}
\caption{Cumulative fraction of the total normalized payoff produced in the network as a function of the nodes ranked from the most productive to the less productive ones for several values of $\alpha$. The colored regions depict the $80\%$ of the total wealth. For the static and linear resource allocation cases ($\alpha = 0.0 $ and $1.0$) more than the $60\%$ of the nodes is needed to produce the $80\%$ of the total wealth while for non linear resource allocations almost the $20\%$ of the nodes produces alone the $80\%$ of the wealth, resembling a {\it Pareto law}\cite{pareto}. Results represent the average over at least $500$ different initial conditions. The other parameters are the same as in Fig.~\ref{fig2}.}
\label{fig4}
\end{figure}

The colored area in Fig.~\ref{fig4} indicates the $80\%$  of the cumulative fraction of the total normalized payoff produced by nodes ranked from the most to the least productive ones. In the classical PGG almost the $70\%$ of the nodes are required to reach the $80\%$ of the wealth while for a quadratic ($\alpha=2.0$) resource allocation setup, this value reaches approximatively the $24\%$ and for higher $\alpha$ becomes more stable and asymptotically approaches the $20\%$ (see SI). Given that we have not imposed any rule on the PGG other than a stochastic investment mechanism and a replicator-like evolution of the strategies,, it can be said that the resulting Pareto Law is obtained from first principles.

\subsection{Negative links} 
The uneven investment distribution observed in Fig.~\ref{fig2} also has another important implication for the games' dynamics. In fact, we noted that if the contribution of a player $i$ in a game is significantly smaller than the average of the other ones, the payoff obtained by the other players is smaller than what they would obtain if player $i$ did not participate, i.e., if the link between $i$ and the focal player of the game doesn't exist.  In this context it is important to notice that, even if player $i$ is formally a cooperator, for that specific game she is acting as a sort of defector as her presence has the effect of reducing the  income of other players. 
Mathematically this condition can be represented by the following inequality:
\begin{equation}
\frac{r}{k_j+1}\left(\sum_{l\in \nu_j}I_{l,j}(t)s_l(t)\right) < \frac{r}{k_j}\left(\sum_{l\in \nu_j\setminus i}I_{l,j}(t)s_l(t)\right)
\label{eq:eq1}
\end{equation}
where $\nu_j$ represents  all the neighbors of agent $j$ while $\nu_j\setminus i$ stands for the same set excluding player $i$.
An important implication of Eq.~(\ref{eq:eq1}) is that it allows to classify every directed link according to whether it is verified or not. In this way we can define a link as {\em positive}  if  eq.~\ref{eq:eq1} is not satisfied -- the contribution on the link is large enough to create an added value in the game -- or, in the opposite way, a link as {\em negative} if the contribution of player $i$ is small enough to satisfy eq.~\ref{eq:eq1} implying that the absence of the link would be beneficial for the other players of the game.

\begin{figure*}
\includegraphics[width=0.8\textwidth]{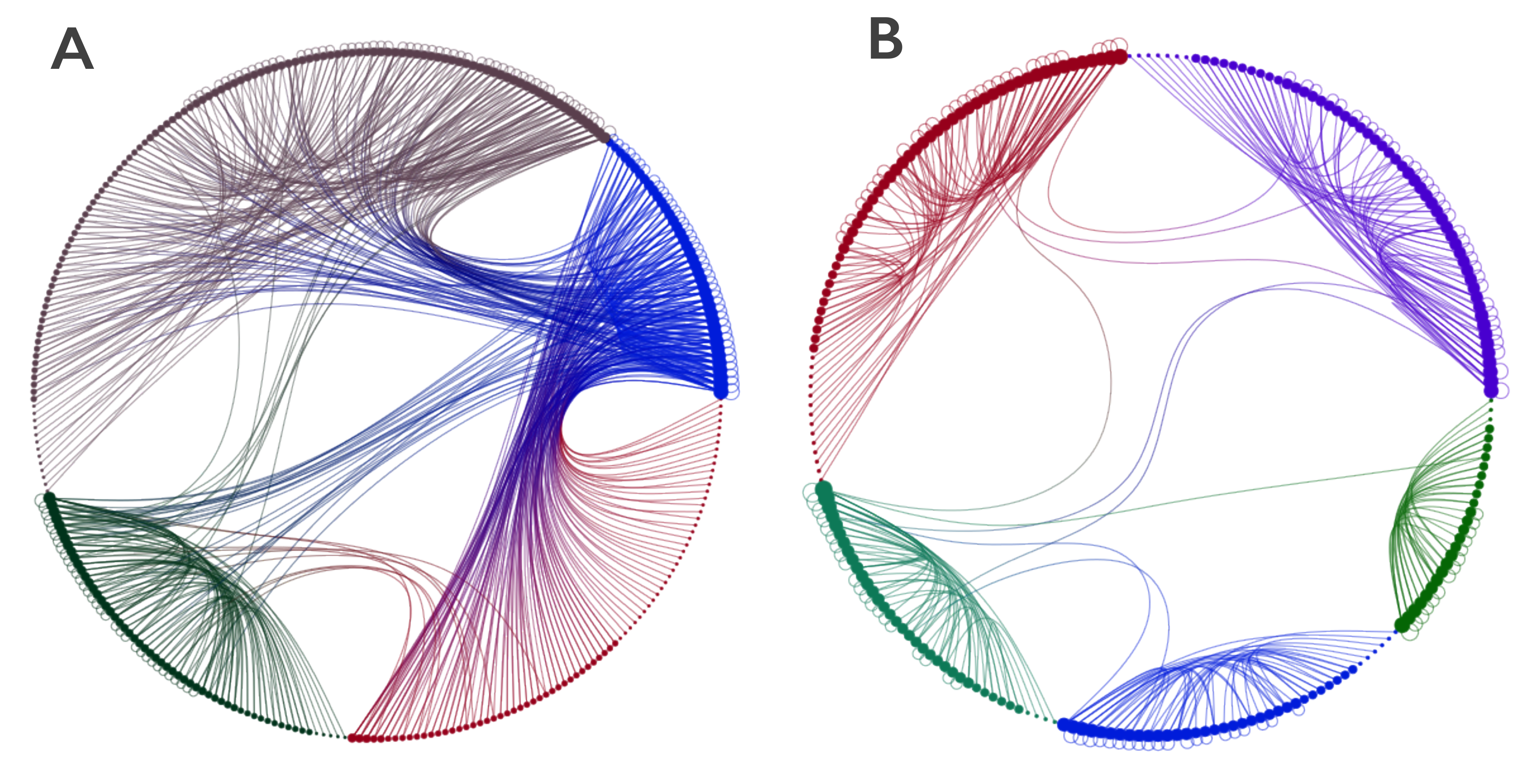}
\caption{Visualization of the networks obtained considering only {\em negative} (panel a) and {\em positive} (panel b) links from the same original network of size $N=300$.  The original graph is a scale-free network generated according to the uncorrelated configuration model\cite{Catanzaro2005} with $\gamma = 2.5$ and $N= 300$ nodes.}
\label{fig5}
\end{figure*}

By analyzing how the two types of links are organized we can dissect the entire network in two subgraphs: one formed only by negative links and the other containing the positive ones. The analysis of the two networks brings about unexpected results. We found that in almost all the realizations the two networks were connected graphs (only in few cases the positive network presented some isolated nodes) and, more importantly, in all the cases the positive network had a backbone-like structure with similar topological features of the minimum spanning tree of the original network. On the other hand, the negative network always includes the majority of the links and its structure strictly resemble the original one (the details of the topological analysis and the comparison between the positive network and the minimum spanning tree are given in the  SI). Also visually (Fig.~\ref{fig5}) the difference between the two networks are notable with the positive network formed by long chains of poorly connected nodes resembling the spanning tree. It is worth stressing that this backbone organization of the links at the entire network level spontaneously emerges as a consequence of the self-organization of the players at the local level without any control mechanism.

Finally, it is also noteworthy that even if all the presented results have been obtained in the so-called \textit{fixed cost per player} paradigm where each individual has the same capital $c$, our results qualitatively hold also for the opposite case of a \textit{fixed cost per interaction} paradigm, where players have a capital $c$ for each game (link) in which they participate (see SI). In addition, to further prove the robustness of our findings we also test different evolutionary rules beyond the finite size equivalent of the replicator dynamics like unconditional imitation, the Fermi rule and a Moran process. In all the cases (not shown) the results reveal the same qualitative behavior and very small quantitative differences. 

\section{Discussion} 
%% \subsubsection{}
Even though heterogeneity has been recognized as one of the most effective mechanisms to favor cooperation in evolutionary games\cite{peng10,lei10,zhang10,yang12,perc07,shi10,zhang_10,gao10,vukov11,cao_10,zhang_12,kun13} some of its consequences  still remain uncovered.

Aimed at shedding light on the organization of cooperation in public goods games, in this paper we have focused on a different rule for  investments that allows players to allocate their resources unevenly. Although this modification might appear not significant, it leads to a radical paradigm shift that can help us explain social and economic hierarchies observed in our complex society. Specifically, despite of its simplicity, our model offers a first-principled explanation of the Pareto Law for wealth distribution\cite{pareto}. This result is a direct consequence of the bimodal distribution of investments observed in Fig~\ref{fig2}, resulting from a behavior in which players invest the majority of their contribution in one game creating small productive clusters of nodes. Although the emergence of those clusters is responsible for the heterogeneity observed in the cumulative payoff distribution (Fig.~\ref{fig4}) the reason why this $80-20$ rule is so stable for a large range of values of $\alpha$ are still unclear and surely deserve further studies.

The heterogenous payoff distribution  is not the only consequence of the uneven investment allocation. Analyzing how players distribute their contribution we also found that if the investment on one link is below the threshold given by Eq.~(\ref{eq:eq1}), the presence of the link is detrimental for the other players leading to a lower payoff. This result imposes a change in how we think about cooperation in evolutionary game theory as one player can act as a cooperator in one game (link) and as a defector in others. This concurrent behavior makes it more meaningful to speak in terms of cooperation in games rather than of cooperators. That is, it suggests that in some contexts, the term cooperation should be carefully used, and that it may be more natural to change the reference of cooperation from players to games. 

It is also important to note that the presence of these so-called \textit{negative} links is not an exclusive feature of our model but it represents a usual situation also in the classical formulation of PGG on heterogeneous networks with the \textit{fixed cost per player} paradigm. To clarify this point, it is instructive to focus on a toy example.  Let's consider a simple network composed by a ring of $n$ nodes in which each node is connected only to its two nearest neighbors and to a hub placed at the center of the ring, i.e., a wheel-like configuration. In this case, the hub will have degree $n$ while the other nodes in the ring degree $3$. In the fixed cost per player setup, the hub will be involved in $n+1$ games and its contribution in each game will be $c/(n+1)$ while the other nodes will participate in $4$ different games and contribute to each $c/(3+1)$.  It is straightforward to demonstrate that in the games centered on the the leaves for $n > 4$ Eq.~\ref{eq:eq1} holds for all  possible values of $r$ and $c$. In this case even if the hub is formally a cooperator its presence reduces the payoff obtained by all the other nodes in the ring.

Eq.~\ref{eq:eq1} allowed us also to classify contributions  as \textit{negative} or \textit{positive} and to split the original network into two layers each one made up of links of the same type. Astonishingly, we found that not only players self-organize their positive links to create highly productive groups but also at a higher level they tend to form a backbone of the entire network. The structure we found strictly resembles the minimum spanning tree of the original interaction network and in most of the cases cover more than $90\%$ of the agents in the system.

Finally, taken together, our findings not only explain the increase in cooperation observed in previous studies\cite{gao10,vukov11,cao_10,zhang_12,kun13} but also can help to understand the basis behind the heterogeneity in wealth distribution observed in almost all human societies and give a hint about the organization  and functioning of large economic systems. Our results also impose a radical change in our idea of cooperation in evolutionary game theory as they demonstrate that also in the classical formulation of Public Goods Games  on networks players can act as cooperators and defectors at the same time. 
%Thus, we think that our work paves the way for a new class of theoretical models.

%% == end of paper:

%% Optional Materials and Methods Section
%% The Materials and Methods section header will be added automatically.

%% Enter any subheads and the Materials and Methods text below.
\section{Methods}
\label{sec:met}
% Materials text
\label{model} We consider a PGG game on networks\cite{santos08} with a dynamical resource allocation scheme that allows individuals to invest higher quantities in profitable groups and reserve their resources from unfavorable ones. Each  node  $i$ of the network is considered as a player participating in $k_{i}+1$ different PGGs with its neighbors. Participating in a PGG round each individual can decide to contribute (cooperate) a part of its resources or act as free-riders (defect). For the fixed cost per player scheme the total amount of resources for each round is fixed to $c$ and equal for all the players, whereas in the fixed cost per game each player has a quantity $(k_{i}+1)c$. In case of cooperation, the contribution of each agent in a game is calculated dynamically and depends on the payoff obtained by the player in the previous round of the game. Specifically, the investment of player $i$ at time $t+1$ in the game where node $j$ is the focal player is defined as $I_{i,j}(t+1)$ and for the fixed cost per player reads as: 
\begin{equation}\label{eq:invest}
I_{i, j}(t+1)=\frac{ce^{\alpha \Pi_{i,j}(t)}}{\sum_{l\in
\nu_i} e^{\alpha \Pi_{i,l}(t)}},
\end{equation}
while in the fixed cost per game: 
\begin{equation}\label{eq:invest2}
I_{i, j}(t+1)=(k_i+1)\frac{ce^{\alpha \Pi_{i,j}(t)}}{\sum_{l\in
\nu_i} e^{\alpha \Pi_{i,l}(t)}},
\end{equation}
where $ \Pi_{i,j}(t)$ is the payoff obtained by agent $i$ in the game centered on node $j$ at the previous time step,  $\nu_i$ is the set of all the first neighbors of node $i$ and $\alpha$ is a parameter that allows to differentiate between a static and homogeneous resource allocation ($\alpha =0$) and heterogeneous distributions where higher resources are invested in best performing games ($\alpha > 0$). At time $t =0$, as all the previous payoffs are set to zero, the contribution is divided evenly between all the games. 
 
In this setting the payoff  $ \Pi_{i,j}(t)$ of player $i$ in the game where $j$ is the focal player can be calculated as: 
\begin{equation}\label{eq:payoff_i}
\Pi_{i, j}(t)=\frac{r}{k_j+1}\left(\sum_{l\in \nu_j}I_{l,j}(t)s_l(t)\right) -I_{i,j}(t)s_i(t),
\end{equation}
where $r$ is the synergy factor and $s_x(t)$  is a dichotomous variable representing cooperation  $s_x(t)=1$ and defection $s_x(t)=0$ respectively.
Summing over all the games in which player $i$ participates the total payoff earned by $i$ at time $t$ reads as: 
\begin{equation}\label{eq:payoff_tot}
\Pi_i(t)=\sum_{j\in \nu_i}\Pi_{i, j}(t).
\end{equation}

At the end of each round  players update synchronously  their strategies according to the finite population equivalent of the replicator dynamics. Each player $i$ selects with uniform probability one of her neighbors $j$ and compare their respective payoffs. If $\Pi_i(t) \ge \Pi_j(t)$ the player will keep its strategy in the next time step otherwise, with probability $P(i \rightarrow j)$ player $i$  will copy the strategy of $j$. We can calculate $P(i \rightarrow j)$ as: 
\begin{equation}\label{eq:repdyn}
P(i\rightarrow j)=\frac{\Pi_j(t)-\Pi_i(t)}{M},
\end{equation}
where $M$ is a normalization factor defined as the  maximum possible payoff difference between two players in the network assuring that $ 0 \le P(i \rightarrow j) \le 1$.

\subsection{Numerical setup}
\label{numsim} Numerical results presented in the text are the average of at least $500$ independent runs with randomly chosen initial conditions. At the beginning of each run players are assigned randomly  one of the two available strategies (cooperate or defect) with probability $0.5$. The average density of cooperators and the other quantities considered  are evaluated at the stationary state after a sufficiently long relaxation time (usually $5\cdot10^4$ time-steps) and then averaged over additional $10^3$ steps. As a substrate we employ scale-free networks generated according to the uncorrelated configuration model \cite{Catanzaro2005,boccaletti06} with mean degree $\langle k \rangle \approx 4, 6, 8$, exponents $\gamma= 2.2, 2.5$ and $2.7$ and sizes $N= 10^3, 5 \cdot 10^3$ and $10^4$. The results presented  are independent of the system size, $\gamma$ and mean degree.
%\end{Methods}

%% Optional Appendix or Appendices
%% \appendix Appendix text...
%% or, for appendix with title, use square brackets:
%% \appendix[Appendix Title]

\begin{acknowledgments}
This work has been partially supported by MINECO through Grants FIS2014-55867-P; Comunidad de Arag\'on (Spain) through a grant to the group FENOL and by the EC FET-Proactive Project MULTIPLEX (grant 317532 to YM and SM). SM is supported by the MINECO through the Juan de la Cierva Program. CYX has been supported by the National Natural Science Foundation of China (NSFC) through Grant No. 61374169
\end{acknowledgments}

%% PNAS does not support submission of supporting .tex files such as BibTeX.
%% Instead all references must be included in the article .tex document. 
%% If you currently use BibTeX, your bibliography is formed because the 
%% command \verb+\bibliography{}+ brings the <filename>.bbl file into your
%% .tex document. To conform to PNAS requirements, copy the reference listings
%% from your .bbl file and add them to the article .tex file, using the
%% bibliography environment described above.  

%%  Contact pnas@nas.edu if you need assistance with your
%%  bibliography.

% Sample bibliography item in PNAS format:
%% \bibitem{in-text reference} comma-separated author names up to 5,
%% for more than 5 authors use first author last name et al. (year published)
%% article title  {\it Journal Name} volume #: start page-end page.
%% ie,
% \bibitem{Neuhaus} Neuhaus J-M, Sitcher L, Meins F, Jr, Boller T (1991) 
% A short C-terminal sequence is necessary and sufficient for the
% targeting of chitinases to the plant vacuole. 
% {\it Proc Natl Acad Sci USA} 88:10362-10366.

%% Enter the largest bibliography number in the facing curly brackets
%% following \begin{thebibliography}

%\end{article}
%%%%%%%%%%%%%%%%%%%%%%%%%%%%%%%%%%%%%%%%%%%%%%%%%%%%%%%%%%%%%%%%

%% Adding Figure and Table References
%% Be sure to add figures and tables after \end{article}
%% and before \end{document}

%% For figures, put the caption below the illustration.
%%

\end{document}